# Brief Report

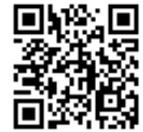

# Optogenetic control of genetically-targeted pyramidal neuron activity in prefrontal cortex

Michael V. Baratta[1], Shinya Nakamura[1], Peter Dobelis[1], Matthew B. Pomrenze[1], Samuel D. Dolzani[1] & Donald C. Cooper[1]*

A salient feature of prefrontal cortex organization is the vast diversity of cell types that support the temporal integration of events required for sculpting future responses. A major obstacle in understanding the routing of information among prefrontal neuronal subtypes is the inability to manipulate the electrical activity of genetically defined cell types over behaviorally relevant timescales and activity patterns. To address these constraints, we present here a simple approach for selective activation and silencing of specific populations of prefrontal excitatory neurons in both *in vitro* and *in vivo* preparations. Rat prelimbic pyramidal neurons were genetically targeted to express a light-activated nonselective cation channel, channelrhodopsin-2, or a light-driven inward chloride pump, halorhodopsin, which enabled them to be rapidly and reversibly activated or inhibited by pulses of light. These light-responsive tools provide a spatially and temporally precise means of studying how different cell types contribute to information processing in cortical circuits. Our customized optrodes and optical commutators for *in vivo* recording allow for efficient light delivery and recording and can be requested at www.neuro-cloud.net/nature-precedings/baratta.

The ability to activate or silence a specific cell type within a neural circuit in a temporally precise fashion is critical for understanding how the prefrontal cortex (PFC) processes the different types of information underlying emotion and cognition. Experimental control over PFC activity has largely relied on traditional neuroscience loss-/gain-of-function tools (e.g., electrical stimulation, pharmacological modulation, surgical ablation) that do provide the required resolution for controlling specific populations of PFC neurons, either on a temporal or spatial scale. Although a temporally fast tool, electrical stimulation, even within a small volume of neural tissue, indiscriminately targets multiple classes of cells, fibers of passage, and terminals within the electrical field. In addition, the use of electrical stimulation does not allow for loss-of-function behavioral experiments as inhibition is not readily possible with this technique. Pharmacological manipulations, as well, lack cell-type specificity and may exert their effects on a timescale that does reflect behaviorally-relevant neuronal activity. Given that almost all cells in the PFC have some degree of overlapping electrical and pharmacological response profiles, an ideal tool for regulating neuronal activity would utilize a physical stimulus that does not affect native neuronal function. Light is an ideal stimulus because most neurons do not express photoreceptors and many of its parameters (wavelength, intensity, temporal pattern, duration) can be brought under experimental control.

The recent development of optogenetic tools provides an exciting prospect for studying the complex and diverse functions of the PFC by enabling bidirectional control over the electrical activity of genetically-targeted cell populations with the use of light. Although it has been known for decades that certain microbial proteins react to light with ion fluxes, only recently have neuroscientists used genetic and viral manipulations to express these light-sensitive membrane proteins for photocontrol of specific groups of neurons. We present here *in vivo* recordings from excitatory prelimbic cortex (PL) neurons transduced with either the light-activated cation channel, channelrhodopsin-2 (ChR2), or the light-driven chloride pump, halorhodopsin (NpHR) in rat.

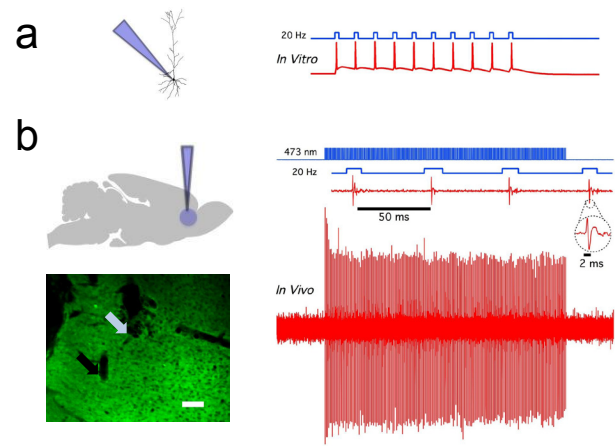

**Fig 1.** Channelrhodopsin-2 (ChR2) induces temporally precise activity in rat prelimbic (PL) cortex. **a)** In vitro schematic (**left**) showing blue light delivery and whole-cell patch-clamp recording of light-evoked activity from a fluorescent CaMKIIα::ChR2-EYFP expressing PL pyramidal neuron (**right**) in an acute brain slice. **b)** *In vivo* schematic (**left**) showing blue light (473 nm) delivery and single-unit recording. (**bottom left**) Coronal brain slice showing expression of CaMKIIα::ChR2-EYFP in the PL. Light blue arrow shows tip of the optical fiber; black arrow shows tip of the recording electrode (**left**). White bar, 100 microns. (**bottom right**) *In vivo* light recording of PL neuron in a transduced CaMKIIα::ChR2-EYFP rat showing light-evoked spiking to 20 Hz delivery of blue light pulses (**right**). Inset, representative light-evoked single-unit response.

---

[1] Department of Psychology and Neuroscience, Institute for Behavioral Genetics, University of Colorado, Boulder, CO 80303 USA
*Correspondence should be sent to D.Cooper@Colorado.edu

                                               

## RESULTS

Before recording from intact rodents, it was critical to confirm that functional ChR2 could be expressed and in the targeted PL excitatory neurons. An adeno-associated viral (AAV) vector carrying the fusion protein ChR2-enhanced yellow fluorescent protein (ChR2-EYFP) under the control of the excitatory neuron-specific promoter CaMKllα. CaMKllα::ChR2-EYFP was stereotactically delivered to the rat PL, where brain slices of the transduced region were made 3 weeks later for whole-cell patch clamp experimentation. Optical stimuli were delivered from a multimode optical fiber (200 μm core diameter, 0.48 NA; Thorlabs) coupled to a 100 mW blue laser. The laser light was positioned directly above the recording site and maximal light intensity failed to produce any detectable postsynaptic responses in nontransduced tissue (data not shown). Under whole-cell current clamp, ChR2-expressing PL neurons exhibited high fidelity spiking in response to illumination with blue light pulses (excitation wavelength, λex, was 473 nm). It should be noted that ChR2-expressing neurons (n=6) were able to reliably follow photostimulation trains up to 20 Hz with 100% fidelity (**Fig. 1a**). This was true for both blue (473 nm) and green (532 nm) laser light pulses.

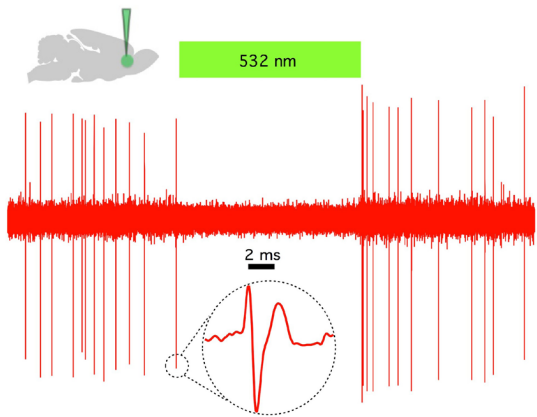

**Fig 2.** Halorhodopsin (NpHR) rapidly and reversibly silences spontaneous activity *in vivo* in rat prelimbic (PL) cortex. (**Top left**) Schematic showing *in vivo* green (532 nm) light delivery and single-unit recording of a spontaneously active CaMKllα::eNpHR3.0- EYFP expressing PL pyramidal neuron. (**Bottom**) Example trace showing that continuous 532 nm illumination inhibits single-unit activity *in vivo*. Inset, representative single unit event; Green bar, (532 nM) is 10 s. *Note silenced responses similar to that depicted in **Fig 2** were observed throughout the dorsal/ventral extent of the PFC (n=6).

For *in vivo* recording, a tungsten electrode (~1.5 MΩ) attached to an optical fiber (200 μm core diameter, 0.48 NA; center-to-center distance between electrode tip and optical fiber tip ~500 μm) that was coupled to either a blue (473 nm) or green (532 nm) diode laser was lowered into the PL to record single unit (n=6) activity in a head-fixed preparation. The light intensity exiting the fiber was approximately ~200 mW/mm$^2$. As with the *in vitro* recordings, *in vivo* delivery of blue light-pulse trains into the PL generated single-units that were capable of following trains of 20 Hz with perfect fidelity (**Fig. 1b**).

We next explored *in vivo* photoinhibition of spontaneous activity in rat PL using an enhanced version of the light-gated third-generation chloride pump NpHR under the CaMKllα promoter (CaMKllα::eNpHR3.0-EYFP). In contrast to what was observed with ChR2, NpHR mediated complete silencing of spontaneous PL activity that was time-locked to the continuous light delivery (λex = 532 nm; **Fig. 2**).

## DISCUSSION

The application of optogenetics opens up an exciting prospect for fine-scale functional analysis of rodent PFC circuitry in which the experimental manipulation is commensurate with the read-out measure (e.g., *in vivo* electrophysiology). Because these photosensitive proteins are genetically targetable, these tools can be used to study the causal function of a variety of defined cell types distributed within intact heterogenous tissue such as the PFC.

## METHODS

All procedures were performed with the approval of the University of Colorado, Boulder Institutional Animal Care and Use Committee. For information on the adeno-associated viral (AAV) vectors used in this study visit www.neuro-cloud.net/nature-precedings/baratta. The final viral concentration was 3 X 10$^{12}$ infectious particles/mL. Viruses were delivered to adult Sprague-Dawley rat PL using a 10 μl syringe and a thin 33 gauge metal needle with a beveled tip (Hamilton Company). The injection volume (1 μl) and flow rate (0.1 μl/min) were controlled with a microinjection pump (UMP3-1; World Precision Instruments). Following injection, the needle was left in place for an additional 10 minutes to allow for virus diffusion. For *in vivo* recording, a tungsten electrode (~1.5 MΩ) attached to an optical fiber (200 μm core diameter, 0.48 NA; center-to-center distance between electrode tip and optical fiber tip ~500 μm) was coupled to either a blue (473 nm) or green (532 nm) diode laser. Custom optrodes and laser couplings can be requested www.neuro-cloud.net/nature-precedings/baratta. For expanded methods visit www.neuro-cloud.net/nature-precedings/baratta.

**PROGRESS AND COLLABORATIONS**
To see up-to-date progress or if you are interested in collaborating with us visit www.neuro-cloud.net/nature-precedings/baratta

**ACKNOWLEDGEMENTS**
This work was supported by National Institute on Drug Abuse grant R01-DA24040 (to D.C.C.), NIDA K award K-01DA017750 (to D.C.C.)

**AUTHOR CONTRIBUTIONS**
MVB, SN, PD, MBP, SDD, DCC performed the experiments, DCC, MVB, SN, PD designed the experiments, DCC and MVB wrote the manuscript.